\documentclass[letterpaper,aps,pre,twocolumn,showpacs,preprintnumbers,superscriptaddress]{revtex4}

\usepackage{graphicx}
\usepackage{dcolumn}
\usepackage{bm}
\usepackage{mathtools}
\usepackage{epsfig}
\usepackage{color}

\begin{document}

\title{Strong anisotropy in two-dimensional surfaces with generic scale invariance: Gaussian and related models}

\author{Edoardo Vivo}
\affiliation{Departamento de
Matem\'{a}ticas and Grupo Interdisciplinar de Sistemas Complejos
(GISC), Universidad Carlos III de Madrid, Avenida de la
Universidad 30, E-28911 Legan\'{e}s, Spain}
\author{Matteo Nicoli}
\affiliation{Physique de la Mati\`ere Condens\'ee, \'Ecole Polytechnique - CNRS,
91128 Palaiseau, France}
\author{Rodolfo Cuerno}
\affiliation{Departamento de
Matem\'{a}ticas and Grupo Interdisciplinar de Sistemas Complejos
(GISC), Universidad Carlos III de Madrid, Avenida de la
Universidad 30, E-28911 Legan\'{e}s, Spain}

\date{\today}

\begin{abstract}
Among systems that display generic scale invariance, those whose asymptotic properties are anisotropic in space (strong anisotropy, SA) have received  relatively less attention, especially in the context of kinetic roughening for two-dimensional surfaces. This is in contrast with their experimental ubiquity, e.g., in the context of thin-film production by diverse techniques.
Based on exact results for integrable (linear) cases, here we formulate a SA Ansatz that, albeit equivalent to existing ones borrowed from equilibrium critical phenomena, is more naturally adapted to the type of observables that are measured in experiments on the dynamics of thin films, such as one and two-dimensional height structure factors. We test our Ansatz on a paradigmatic nonlinear stochastic equation displaying strong anisotropy like the Hwa-Kardar equation [Phys.\ Rev.\ Lett.\ {\bf 62}, 1813 (1989)], which was initially proposed to describe the interface dynamics of running sand piles. A very important role to elucidate its SA properties is played by an accurate (Gaussian) approximation through a nonlocal linear equation that shares the same asymptotic properties.
\end{abstract}


\maketitle

%

\section{Introduction}

Generic Scale Invariance (GSI) broadly makes reference to the appearance of (quasi-) long-range order
in sizable regions of the phase diagram of a given system. It can occur both in quantum and in classical
systems and in equilibrium and out of equilibrium \cite{belitz:2005}. In the case of classical, far-from-equilibrium
phenomena, it provides a natural framework for, among other, conspicuous systems such as those exhibiting
self-organized-criticality (SOC) \cite{grinstein:1990,grinstein:1991,grinstein:1995,christensen:book}. Thus, it embodies the behavior that was naturally attached to systems such as model sand piles, which were thought to reach a critical point
without the need to fine-tune a control parameter. Although the mechanism for criticality of many SOC systems
has turned out to  be more elaborate than this \cite{dickman:2000}, the GSI idea continues to play an important
reference for non-equilibrium systems. Other specific contexts where it is at play are driven diffusive
systems \cite{schmittmann:book} and surface kinetic roughening \cite{barabasi:book}.

Indeed, in terms of their natural description through Langevin equations with time-dependent noise \cite{kardar:book},  the two main dynamical classes of kinetically rough surfaces (see, e.g., Ref. \cite{cuerno:2004} and references therein) necessarily display GSI \cite{grinstein:1991,grinstein:1995}; namely systems with nonconserved dynamics whose evolution equation is invariant under arbitrary shifts, $h \to h + \mbox{const}$, in the values of the surface height $h$ and systems with conserved dynamics and non-conserved noise. We will henceforth focus on these two groups of systems. A very important physical difference with usual SOC systems concerns the relevant time scales \cite{grinstein:1991,grinstein:1995}. Thus, while in SOC systems there is a wide time-scale separation between driving (infinitely slow) and system response (infinitely fast), this is {\em not} the case for kinetically rough surfaces, for which noise is usually taken as time dependent, akin to thermal noise \cite{barabasi:book,
kardar:book}.

In the kinetic roughening context, to date most experimental and theoretical work has been devoted to characterize and understand scaling behavior that is isotropic in space. This is partly due to the fact that even the most natural anisotropic generalizations of the paradigmatic Kardar-Parisi-Zhang (KPZ) and conserved KPZ equations, considered in Refs.\ \cite{wolf:1991} and \cite{kallabis:1998}, respectively, turn out to display isotropic behavior themselves. However, to some extent, this should strike us as unexpected. Considering, for instance, the application of kinetic roughening to standard contexts, like epitaxial growth of thin films \cite{misbah:2010}, or crystal growth from a melt \cite{davis:book}, physical anisotropies are ubiquitous in, e.g., energetic barriers for surfaces to relax by surface diffusion, or in surface tension effects. Thus, it would be natural to expect the occurrence of anisotropic scaling properties in the asymptotic states of (some of) such systems.

For instance, for thin films grown by Molecular Beam Epitaxy (MBE) on surfaces that are vicinal to a high-symmetry orientation, the direction of the average surface tilt and the average orientation of the ensuing steps play very different physical roles, frequently leading to anisotropic behavior as for the step bunching and/or meandering instabilities in Si(001) \cite{schelling:1999,uwaha:1999,verga:2009_12}. Even under morphologically
stable MBE growth conditions, spatial anisotropies may occur, as for growth of GaAs films \cite{ballestad:2001,ballestad:2002}. Still with nanoscopic systems, erosion, rather than growth, of thin films by
ion-beam sputtering (IBS) induces space anisotropies related with the different roles played by the direction on the target that lies along the projection of the ion beam and the direction perpendicular to it \cite{chan:2007,munoz-garcia:2009}. Reaching the realm of macroscopic spatially extended systems, fracture of solids provides still another instance for
the occurrence of space anisotropies, in this case between the crack propagation and crack front directions \cite{alava:2006,bonamy:2011}.

Attempts have been made to explore the general role of anisotropies in kinetic roughening (see, e.g., Refs.\ \cite{tauber:2002,tauber:book} and references therein), and some observations and characterizations of anisotropic behavior are available \cite{zhao:book}. However, for the physical case of two-dimensional surfaces, general conditions are not known on the occurrence of isotropic vs anisotropic behavior, and neither is a formulation (Ansatz) of the latter available, that can be readily applied to analyze experimental data on, say, surface dynamics of thin films. Indeed, anisotropic kinetic roughening is encoded into a scaling Ansatz \cite{schmittmann:2006} that originates in the study of critical dynamics of equilibrium statistical-mechanical systems \cite{henkel:book_v1}. Such a formulation is quite powerful from the theoretical
point of view (enabling analysis of scaling properties for arbitrary substrate dimension, etc.) but is not
particularly natural for the characterization of actual two-dimensional surfaces. In this way, the experimental community does not have a clear guiding principle that allows us to identify correctly the behavior that should be expected for each observable under conditions for strong anisotropy, and comparison with theoretical models is thus hampered.
Alternative formulations are available \cite{zhao:book} but have been tested onto linear models only \cite{zhao:1998}, so  their generality has not been checked. Nor has a systematic study been done on the relation between different observables (like height correlation functions or 1D and 2D height structure factors). 
Therefore, a fully consistent experimental analysis and a quantitative comparison with theoretical models cannot be done unless such relationship is fully clarified.

The aim of this paper is to put forward a formulation of the dynamic scaling Ansatz for strongly anisotropic kinetically
rough surfaces that is directly adapted to the analysis of experimental and/or simulation data for the physical case of two-dimensional interfaces, while being founded on the behavior displayed by model systems. To this end, we consider interface equations that can either be solved or that can be mapped into solvable cases. In general, we expect
that progress achieved can allow a better assessment of the experimental impact of strongly anisotropic kinetic roughening, and, thus, add to an improved understanding of space anisotropies in large classes of GSI systems, for the physical two-dimensional case.

In Sec.\ \ref{sec_ansatz} of the paper, we, first, formulate our Ansatz and explore its implications for the various observables that are usually employed, both in experiments and in numerical studies of discrete or continuum models. The motivation for the Ansatz that is based on exact results for linear equations is discussed in Sec.\ \ref{sec_lin}.
Actually, recent experimental validation has been achieved in the interpretation of experiments of surface erosion by ion-beam sputtering \cite{us_exp}. Then, in Sec.\ \ref{sec_hk}, we again verify the Ansatz against a nonlinear equation that has played a relevant role in the context of space anisotropies. As a paradigmatic system, we specifically consider the Hwa-Kardar (HK) equation that describes the dynamics of a running sand pile \cite{hk,kardar:book} and is a particular instance of conserved interface dynamics with nonconserved noise. Through direct numerical simulations, we analyze in detail its anisotropic roughening behavior in the light of our scaling Ansatz. Some apparent inconsistencies (subleading corrections) are solved thanks to the mapping of the HK equation to a nonlocal linear equation that shares the  same asymptotic properties (as conjectured \cite{hk}). A discussion of these results is provided in Sec.\ \ref{sec_disc}.

\section{Anisotropic scaling Ansatz}
\label{sec_ansatz}

For a two-dimensional rough interface, a straightforward generalization of the height-difference correlation
function that is usually considered for isotropic systems \cite{barabasi:book},
\begin{equation}
G(r) = \langle [h(\mathbf{r}+\mathbf{r}_0)-h(\mathbf{r}_0)]^2 \rangle ,
\label{G}
\end{equation}
is provided by analogous correlation functions along each one of the two space directions, namely,
\begin{eqnarray}
&& G_x(x) = \langle \left[h(x_0+x,y_0) - h(x_0,y_0)\right]^2\rangle, \label{eq_dGx} \\[5pt]
&& G_y(y) = \langle \left[h(x_0,y+y_0) - h(x_0,y_0)\right]^2\rangle. \label{eq_dGy}
\end{eqnarray}
In Eqs. \eqref{G} through \eqref{eq_dGy}, $h(\mathbf{r})$ is the surface height above point $\mathbf{r}=(x,y)$ on
a substrate plane where $\mathbf{r}_0=(x_0,y_0)$ is an arbitrary position, and brackets denote averages over the noise distribution (e.g., independent realizations of the experiment). In general, it is the physical conditions that dictate the two appropriate directions that play the role of $x$ and $y$ here. Thus, under IBS at oblique incidence, the projection of the ion beam on the target plane fixes one of them. Likewise, in MBE of vicinal surfaces, it is the direction of the average tilt with respect to the height symmetry orientation that is fixing one main direction, and so on.

While the statement of kinetic roughening in an isotropic system implies the simple power-law behavior $G(r) \sim r^{2 \alpha}$, where $r=|\mathbf{r}|$ and $\alpha$ is the roughness exponent \cite{barabasi:book}, the expected behavior in the presence of anisotropies is, rather \cite{hk},
\begin{equation}
G_x(x) \sim x^{2\alpha_x}, \qquad \label{eq:heightdiff_scaling_x_y}
G_y(y) \sim y^{2\alpha_y}.
\end{equation}
where, in principle, two different roughness exponents exist, $\alpha_x$ and $\alpha_y$. We will speak of strong
anisotropy (SA) if they take different values in the asymptotic state of the system, while this will be referred to as
of weak anisotropy (WA) if they take the same values. Note incidentally that in the present discussion we will assume
that the surface is at a steady state at which statistical properties are time invariant. Later, we consider the time evolution explicitly.

Naturally, real-space correlations are not the only observables one can measure from, e.g., simulations or from actual experimental data. In the context of surface kinetic roughening, a very important quantity is the two-dimensional
power spectral density (PSD) or structure factor of the surface height,
\begin{equation}
S(\mathbf{k}) = \langle |\tilde{h}(\mathbf{k})|^2 \rangle ,
\label{psd}
\end{equation}
where $\tilde{h}(\mathbf{k})$ is the space Fourier transform of $h(\mathbf{r}) - \bar{h}$, with $\bar{h}$ being the space
average of the height. On the one hand, this function can be easily measured in many experimental setups by using, e.g., X-ray diffraction techniques \cite{zhao:book}. On the other hand, for theoretical modeling and simulation, it has many advantages over real-space correlation functions, specially in the presence of crossover behavior and/or anomalous scaling; see, e.g., Ref.\ \cite{cuerno:2004} and references therein. As will be justified in the next sections, we will hypothesize the following long-time, large-distance behavior of the PSD for an anisotropic surface:
\begin{equation}
\label{eq_ASA}
S(\mathbf{k}) = S(k_x,k_y) \sim \frac{1}{k_x^{2\tilde \alpha_x} +\nu k_y^{2\tilde \alpha_y}},
\end{equation}
where the exponents measured in Fourier space $\tilde \alpha_{x,y}$ will be related somehow to the roughness exponents $\alpha_{x,y}$ and $\nu$ is a mere constant. Our aim is precisely to see how the behavior we have just assumed translates into scaling behavior for other observables in the system. Note that Eq.\ \eqref{eq_ASA} is an analog for nonconserved noise fluctuations of the behavior found in driven-diffusive systems \cite{schmittmann:book}.

As mentioned above, frequently, e.g., in mentioned MBE or IBS systems, physical properties and geometric constraints dictate the appropriate choices for the $x$ and $y$ directions. Nevertheless, as will be shown below, under conditions for strong anisotropy any choice of two orthogonal directions will lead to the same set of two different exponents $\tilde{\alpha}_{x,y}$, which guarantees the generality of Ansatz \eqref{eq_ASA}. In the case of fracture experiments, alternative choices for anisotropic scaling Ans\"atze are also available, in which, e.g., either an auxiliary dynamics is postulated \cite{bonamy:2011,ponson:2006} or  expansions of observables over appropriate functional bases are performed that exploit the fact that isotropic materials often have anisotropic fracture surfaces only because of the breaking of isotropy by the initial conditions \cite{bouchbinder:2005,bouchbinder:2006}.

Once a choice of appropriate or convenient $x,y$ directions has been made, two rather natural observables are the PSDs $S_x(k_x)$ and $S_y(k_y)$ of corresponding {\em one-dimensional} profiles. In e.g.\ thin film experiments, such observables are readily measured using scanning probe microscopies (AFM or STM) \cite{zhao:book}. Thus, for instance, considering a fixed value $y=y_0$, one defines
\begin{equation}
S_x(k_x) = \langle \tilde h_{y_0}(k_x) \tilde h_{y_0}(-k_x) \rangle ,
\label{def_Sx}
\end{equation}
where $\tilde h_{y_0}(k_x)$ is the Fourier transform of the corresponding one-dimensional profile $h(x,y_0)$,
\begin{equation}
\label{eq:slice_fourier}
\tilde h_{y_0}(k_x) = \dfrac{1}{\sqrt{2\pi}}\int_{-\infty}^{+\infty} dx\, e^{i x k_x} h(x,y_0).
\end{equation}
Here we have considered the $L\to\infty$ limit of a finite system with lateral size $L$ and periodic boundary conditions. 
Such a limit will be reconsidered when needed below.

It is straightforward to obtain the relation between these observables [the same procedure as applied to $S_x(k_x)$ leads to the definition of $S_y(k_y)$ after exchanging labels $x\leftrightarrow y$] and the two-dimensional PSD. Using
\eqref{eq:slice_fourier} in \eqref{def_Sx}, and substituting $h(x,y_0)$ with its representation in terms of the two-dimensional Fourier transform $\tilde h(k_x, k_y)$,
\begin{equation}\label{eq:slice_fourier2}
\tilde h_{y_0}(k_x) = \dfrac{1}{\sqrt{2\pi}} \int_{-\infty}^{+\infty} dk_y \, e^{-i y_0 k_y}\tilde h(k_x, k_y),
\end{equation}
we obtain
\begin{equation*}
\begin{split}
S_x(k_x) 
& =\dfrac{1}{2\pi} \int_{-\infty}^{+\infty} \hskip - 5pt dk_y  dk'_y\,  e^{-i y_0 (k_y + k'_y)} \langle \tilde h(k_x, k_y) \tilde h(-k_x, k'_y)\rangle \\
&= \dfrac{1}{\pi} \int_0^{\infty}dk_y \, S(k_x,k_y).
\end{split}
\label{eq:derivation_psd1d_psd2d}
\end{equation*}
Following the same steps, we obtain an analogous expression for $S_y(k_y)$; in summary,
\begin{eqnarray}
&& S_x(k_x) = \dfrac{1}{\pi} \int_0^{\infty} dk_y \, S(k_x,k_y),  \label{eq_Sx}\\[5pt]
&& S_y(k_y) = \dfrac{1}{\pi} \int_0^{\infty} dk_x \, S(k_x,k_y). \label{eq_Sy}
\end{eqnarray}
Now, substituting the scaling hypothesis \eqref{eq_ASA} into Eqs.\ \eqref{eq_Sx} and \eqref{eq_Sy}, we
obtain the following scaling laws for the Fourier transform of the one-dimensional cuts:
\begin{eqnarray}
&& S_x(k_x) \sim k_x^{-(2\tilde \alpha_x - \zeta)},  \label{eq_sSx} \\[5pt]
&& S_y(k_y) \sim k_y^{-(2\tilde \alpha_y - 1/\zeta)}, \label{eq_sSy}
\end{eqnarray}
where $\zeta = \tilde{\alpha}_x /\tilde{\alpha}_y$ is the so-called anisotropy exponent. SA holds whenever $\zeta \neq 1$, whereas $\zeta = 1$ denotes WA, meaning that the steady state of the system is actually \emph{isotropic}.
As shown below \cite{keller:2009}, Eqs.\ \eqref{eq_sSx} and \eqref{eq_sSy} are equivalent to
\begin{eqnarray}
\label{eq:psd1d_scaling_b}
&&S_x(k_x)\sim k_x^{-(2\alpha_x+1)} , \\[5pt]
&&S_y(k_y)\sim k_y^{-(2\alpha_y+1)} , \label{eq:psd1d_scaling_b2}
\end{eqnarray}
that provide the natural generalization to the SA case of the scaling behavior of the PSD of 1D cuts of the surface in the isotropic case, in which 
\mbox{$\alpha_x=\alpha_y=\alpha$} and \mbox{$S_{x,y} \sim k_{x,y}^{-(2\alpha+1)}$} \cite{hansen:2001}.

We still have to prove that the anisotropy condition $\zeta\neq 1$ indeed corresponds to $\alpha_x \neq \alpha_y$, for which we need to relate these real-space exponents with their momentum-space counterparts, $\tilde{\alpha}_{x,y}$. Such a
relation is obtained from the definition of $G_{x,y}$, Eqs.\ \eqref{eq_dGx} and \eqref{eq_dGy}.
The Fourier representation of the one-dimensional cut is
\begin{equation}\label{eq:slice_fourier_back}
h(x,y_0) = \dfrac{1}{\sqrt{2\pi}}\int_{-\infty}^{+\infty} dk_x \, \tilde h_{y_0}(k_x)e^{-i k_x \, x},
\end{equation}
where $\tilde h_{y_0}(k_x)$ is given by Eq.\ \eqref{eq:slice_fourier}. Using the latter into \eqref{eq_dGx}, it is easy to see that
\begin{equation}\label{eq:heightdiff_anisotropic_vs_psd1d}
G_x(x) = \dfrac{2}{\pi} \int_0^\infty dk_x  \left[1-\cos(k_x \, x)\right] S_x(k_x),
\end{equation}
a 1D version of the general 2D relation between the height-difference correlation function and the height structure factor
\cite{zhao:book}. Now, using the scaling Ansatz for the one-dimensional PSD, Eqs.\ \eqref{eq_sSx} and \eqref{eq_sSy},
we are able to  write the scaling behavior of
\begin{equation}
\label{eq_saGx}
G_x(x) \sim x^{2\tilde \alpha_x - \zeta -1},
\end{equation}
and
\begin{equation}
\label{eq_saGy}
G_y(y) \sim y^{2\tilde \alpha_y - 1/\zeta -1},
\end{equation}
so 
\begin{eqnarray}
&& 2\alpha_x = 2\tilde \alpha_x - \zeta -1, \label{eq_relalphax} \\[5pt]
&& 2\alpha_y = 2\tilde \alpha_y - 1/\zeta -1.  \label{eq_relalphay}
\end{eqnarray}
Equations \eqref{eq_relalphax} and \eqref{eq_relalphay} lead immediately to Eqs.\ \eqref{eq:psd1d_scaling_b} and \eqref{eq:psd1d_scaling_b2} and  to the fact that a single anisotropy exponent is defined, namely
\begin{equation}
\zeta = \dfrac{\tilde{\alpha}_x}{\tilde{\alpha}_y} =  \dfrac{\alpha_x}{\alpha_y}.
\end{equation}

In order to incorporate the time dependence into our scaling Ansatz, we consider another important observable, namely the global surface roughness. In anisotropic systems, it is natural to measure the roughness of slices of the surface along the $x$ and $y$ axes, $W_x(t)$ and $W_y(t)$, respectively. These are simply obtained as the averages of Eqs.\
\eqref{eq_dGx} and \eqref{eq_dGy} over all possible values of $x_0$ and $y_0$, respectively. The Family-Vicsek (FV)
dynamic scaling Ansatz \cite{barabasi:book} that applies to isotropic kinetic roughening is typically formulated in terms of the short- and long-time behavior for the surface roughness $W^2(t) = \langle (h-\bar{h})^2 \rangle = \int S(\mathbf{k}) \; {\rm d}\mathbf{k}$. Thus \cite{barabasi:book}, $W \sim t^{\beta}$ for $t \ll t^{1/z}$, while $W \sim L^{\alpha}$ for $t \gg t^{1/z}$, where $z$ is an independent exponent, $t^{1/z}$ is proportional to the length-scale below which nontrivial correlations have built up among height values at different substrate positions, and $\beta = \alpha/z$.


In the SA case, one might expect that a similar behavior would hold for the roughness functions
of independent 1D cuts, $W_{x,y}(t)$, but with two different growth exponents, $\beta_x=\alpha_x/z_x$ and $\beta_y=\alpha_y/z_y$. However, it is not difficult to see that these two growth exponents coincide for any value of the anisotropy exponent $\zeta$. To prove this, we consider two different anisotropic scale transformations of the coordinates and the surface height, one along the $x$ coordinate, $\vec{r}_1\equiv (x_1,y_1,t_1,h_1)$, and another one along the $y$ direction, $\vec{r}_2\equiv (x_2,y_2,t_2,h_2)$, with
\begin{eqnarray}
&& \vec{r}_1 = (bx,b^\zeta y, b^{z_x} t, b^{\alpha_x} h), \label{r1} \\[5pt]
&& \vec{r}_2 = (\tilde b^{\tilde \zeta}x,\tilde b y, \tilde b^{z_y} t, \tilde b^{\alpha_y} h), \label{r2}
\end{eqnarray}
where $b$ and $\tilde b$ are arbitrary positive factors. In these expressions, we have incorporated the shape of the different rescaled coordinates under which scale invariance holds. Notice, in principle one has to allow for a ``response'' of the time coordinate to a rescaling in space that is anisotropic, namely $z_x \neq z_y$. Now the hypothesis
of scale invariance implies that the same statistical properties hold after we impose any of the two
rescalings, \eqref{r1} or \eqref{r2} \cite{pastor-satorras:1998}. In particular, we can equate the two forms of rescaling. For the $x$ coordinate, we get $b = \tilde b^{\tilde \zeta}$. We can now apply the same procedure to the $y$
coordinate and, from \mbox{$b^\zeta y  = \tilde b y$}, we get $\tilde\zeta = \zeta^{-1}$. Then, by equating the two different ways of
rescaling the time variable, from $b^{z_x} = \tilde b^{z_y}$  we have $\tilde\zeta = z_y/z_x$.
Finally, from the definition of the growth exponents $\beta_{x,y} = \alpha_{x,y}/z_{x,y}$, we obtain
\begin{equation}
\label{eq_eqbeta}
\frac{\beta_x}{\beta_y} = \frac{\alpha_x z_y}{z_x \alpha_y} = \zeta \, \tilde \zeta  = 1,
\end{equation}
indeed implying $ \beta_x = \beta_y = \beta$ for any value of $\zeta$. Summarizing, there are only three independent critical exponents characterizing a time-dependent SA surface, e.g., $\alpha_x$, $\zeta$, and $z_x$, from which all other exponents described in this section can be obtained. In particular, the behavior of the roughness of 1D line profiles
is $W_{x,y} \sim t^{\beta}$ for $t \ll t^{1/z_{x,y}}$, while $W_{x,y} \sim L^{\alpha_{x,y}}$ for $t \gg t^{1/z_{x,y}}$ with the corresponding values of exponents in terms of the three independent ones. In the next sections we will consider
a number of examples for which we will fully verify all the scaling relations just derived.

\section{Linear equations}
\label{sec_lin}

In this section we demonstrate a direct way of constructing strongly anisotropic systems, by mixing two solvable
one-dimensional equations with different scaling exponents. Thus we are able to combine two distinct
physical phenomena acting along the two directions of the same system. As we will see, the resulting models
will display SA that will be assessed through the scaling Ansatz proposed in the previous section. This will be
done comparing the exact solution with direct numerical simulations, as a prelude to the next section in which
a non-linear equation will be studied.

For simplicity, the equation of motion will be written down for the Fourier components of the surface height.
Thus, we consider
\begin{equation}
\label{eq_nm}
\partial_t \tilde{h}({\bf k},t) = -\left( \nu_x |k_x|^n + \nu_y |k_y|^m\right) \tilde{h}({\bf k},t) + \tilde{\eta}({\bf k},t).
\end{equation}
Here ${\bf k} $ is defined in the two-dimensional space $[-\pi,\pi]^2$ and the term $\tilde{\eta}({\bf k},t)$ is the Fourier transform of a standard time-dependent Gaussian white noise with zero mean and constant variance $2D$. The equation just written actually defines a two-parameter family of models. Each single case corresponds to a specific choice of the pair of positive {\em real} numbers $(n,m)$. In principle, here we only consider positive values for coefficients $\nu_{x,y}$ so  the dispersion relation of Eq.\ \eqref{eq_nm}, namely
\begin{equation}
\sigma(\mathbf{k}) = -\left(\nu_x |k_x|^n + \nu_y |k_y|^m\right),
\label{ec_sigma}
\end{equation}
is negative definite. In this way, Eq.\ \eqref{eq_nm} is morphologically stable. For related morphologically unstable
nonlinear models, see Refs.\ \cite{nicoli:2009a,nicoli:2009b,nicoli:2011}. Note also that, for generic values of $n$, $m$, Eq.\ \eqref{eq_nm} is nonlocal in space, since it is not possible to inverse-Fourier transform it to have a local equation for the surface height $h(x,y,t)$ in real space. This is only possible if $n$, $m$ are even integers. See below for an example.

Equation \eqref{eq_nm} can be explicitly solved for a flat initial condition $\tilde{h}({\bf k},t=0)=0$,
\begin{equation}
\tilde{h}({\bf k},t) = \int_0^t ds \exp\left[\sigma({\bf k})(t-s)\right] \tilde{\eta}_{\bf k}(s).
\label{sol_nm}
\end{equation}
Using this solution, we can compute analytically the two-dimensional PSD for any of these linear models,
\begin{equation}
\label{eq_St}
S({\bf k},t) = \dfrac{D}{2\pi^2} \left(\dfrac{e^{2\sigma({\bf k}) t}-1}{\sigma({\bf k})} \right).
\end{equation}
Due to the morphological stability condition, the exponential term vanishes in the \mbox{$t\to\infty$} limit and Eq.\ \eqref{eq_St} reduces trivially to the scaling Ansatz \eqref{eq_ASA}. For a finite-size system, it is sufficient to verify the condition $t> t_{\rm sat} = \max_{x,y}(L^{1/z_{x,y}})$ in order for the asymptotic form  $S(k_x,k_y)$ to approximate reasonably the full time-dependent $S({\bf k},t)$.

To anticipate the type of analysis that will be performed in the next section for a more complex (nonlinear)
equation, it is instructive to consider here a specific example. Namely we consider a case that combines
evaporation-condensation along the $x$ axis and surface diffusion along the $y$ axis. The resulting equation is a mixture of a 1D Edward-Wilkinson (EW) equation  \cite{barabasi:book} along the former direction and a 1D Linear Molecular Beam Epitaxy (LMBE) equation  \cite{barabasi:book} along the latter, namely
\begin{equation}
\label{eq_2-4}
\partial_t h = \nu \partial_x^2 h - \mathcal{K} \partial_y^4 h + \eta(x,y,t).
\end{equation}
This equation is simply the particular case $(n,m)=(2,4)$ of Eq.\ \eqref{eq_nm} with $\nu_x = \nu$ and $\nu_y={\cal K}$, which can indeed be written down in real space, in terms of local operators acting on $h(x,y,t)$.
We will henceforth call this the \emph{2-4 equation}, to recall the order of derivatives acting along each direction.
Naturally, it is a particular case of a more general linear model in which evaporation-condensation
and surface diffusion are fully anisotropic, namely,
\begin{equation}
\begin{split}
\partial_t h  = &\,   \nu_x \partial^2_x h + \nu_y \partial^2_y h  \\
 & -  \mathcal{K}_{xx}\partial^4_x h - \mathcal{K}_{xy}\partial^2_x \partial^2_y h - \mathcal{K}_{yy} \partial^4_y h + \eta .
\end{split}
\label{eq_2-4gen}
\end{equation}
Equations of this form appear in the context of thin film growth and erosion~\cite{zhao:1998} and have partly been studied
along the lines that follow.

\begin{figure}[t!]
\centering
\epsfig{clip=,width=0.5\textwidth,file=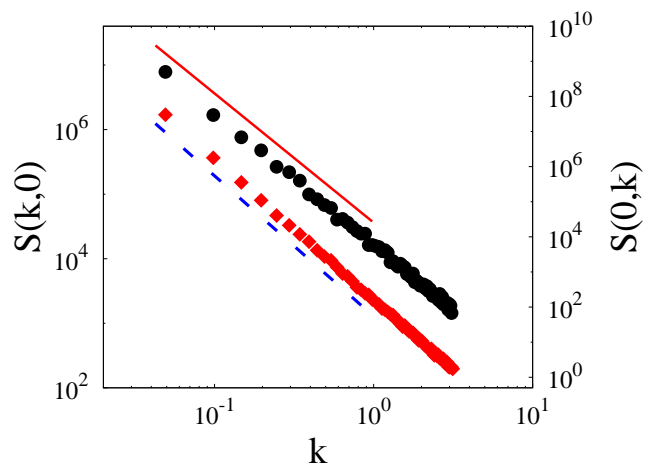}
\caption{(Color online) One-dimensional projections $S(k,0)$ (black circles, left axis) and $S(0,k)$
(red diamonds, right axis) of the two-dimensional PSD of surfaces generated by the 2-4 equation, \eqref{eq_2-4}. Numerical integrations were performed with parameters $\nu = 1$, $\mathcal{K} = 100$, $D=1$, $L=128$, $\Delta x = 1$, and $\Delta t = 1/2$. The saturated two-dimensional PSD was averaged over 150 different noise realizations. The solid red line is a guide for the eye with slope $-2$, while the dashed blue line has slope $-4$. All units are arbitrary.}
\label{fig_2-4PSD}
\end{figure}

Equation \eqref{eq_2-4} being linear, we can indeed obtain all its scaling exponents exactly. Thus, the asymptotic two-dimensional PSD of the 2-4 equation behaves simply as
\begin{equation}
\label{eq_2-4S}
S(k_x,k_y) \sim \left(\nu k_x^2 + \mathcal{K}k_y^4\right)^{-1},
\end{equation}
so  $\tilde \alpha_x = 1$, $\tilde \alpha_y = 2$ and the anisotropic exponent is $\zeta = 1/2$; hence,
the system displays strong anisotropy as expected. The one-dimensional PSD of cuts along the coordinate axes should scale according to Eqs. \eqref{eq_sSx} and \eqref{eq_sSy}, hence,
\begin{equation}
\label{eq_2-4PSD1d}
S_x(k_x) \sim  k_x^{-3/2},\qquad S_y(k_y) \sim  k_y^{-2},
\end{equation}
from which we obtain direction-dependent roughness exponents $\alpha_x = 1/4$ and $\alpha_y = 1/2$. Note \cite{zhao:1998} that these values  differ from those of the EW and the LMBE universality classes in 1 and 2 dimensions \cite{barabasi:book}.

As far as the roughness is concerned, as we have seen before the growth exponent $\beta$ is the same for the two directions. To compute its value, in this case it suffices to rescale Eq.\ \eqref{eq_2-4} according to the transformation $\vec{r}_1$; see Eq.\ \eqref{r1}. Given the existence of the self-affine solution [Eqs.\ \eqref{ec_sigma} and \eqref{sol_nm}], we can impose scale invariance on the rescaled equation,
\begin{equation}
\partial_{t}h=\nu b^{2-z_x}\partial^2_{x}h -\mathcal{K}b^{4\zeta-z_x}\partial^4_{y}h
+ b^{(2\alpha_x-z_x+1+\zeta)/2}\eta ,
\label{ec_resc}
\end{equation}
namely one can request coefficients to be $b$ independent. Thus, substituting the already-known exponents \mbox{$\alpha_x=1/4$} and \mbox{$\zeta=1/2$} we find the dynamic exponents $z_x=2$ and \mbox{$z_y=4$}, so  $\beta = 1/8$.

In order to check these analytical results, we have performed numerical simulations of the 2-4 equation
by means of a pseudospectral integration algorithm as described in Ref.\ \cite{nicoli:2008} and references
therein. First, we have checked the validity of the asymptotic scaling Ansatz \eqref{eq_2-4S} for the two-dimensional PSD.
In Fig.\ \ref{fig_2-4PSD} we present two projections of $S(k_x,k_y)$ along the $k_x$ and $k_y$ axes,
together with the analytical solutions, with very good agreement in the asymptotic regime. Note that, in this and all remaining figures, simulation data will be provided through symbols, while theoretically expected scaling (whether approximate or exact) will be provided as reference lines.

\begin{figure}[t!]
\centering
\epsfig{clip=,width=0.5\textwidth,file=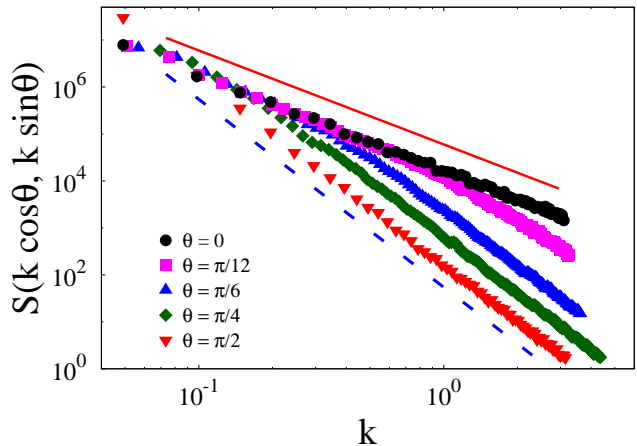}
\caption{(Color online) One-dimensional cuts $S(k \cos \theta, k \sin \theta)$ of the two-dimensional PSD of surfaces generated by the 2-4 equation \eqref{eq_2-4} along oblique directions in $\mathbf{k}$ space, $\theta = 0, \pi/12, 
\pi/6, \pi/4$, and $\pi/2$, represented as functions of $k$. Numerical integrations were performed as in Fig.\ \ref{fig_2-4PSD}. Note the expected scalings $-2$ and $-4$ are best achieved for $\theta$ close
to 0 and $\pi/2$, respectively. All units are arbitrary.}
\label{fig_2-4PSD_add}
\end{figure}

We can, moreover, verify that the same two exponent values $\tilde{\alpha}_{x,y}$ are actually obtained if one performs any
cut of the PSD along any direction in the two-dimensional $\mathbf{k}$ space, see Fig.\ \ref{fig_2-4PSD_add}, the $x$ and $y$ directions being optimal choices with respect to a clear-cut scaling behavior. This guarantees that the strongly anisotropic behavior found is not affected by an arbitrary choice of the $x,y$ directions in an associated physical system. Several different choices are shown in the figure, all of which take the form $S(k \cos \theta, k \sin \theta)$ for different values of the polar angle $\theta$. Thus, e.g., $S(k,0)$, or $S(0,k)$ in Fig.\ \ref{fig_2-4PSD} are just the special $\theta=0$ and $\theta=\pi/2$ cases, respectively. For other choices of $\theta$, the scaling behavior of the corresponding cut $S(k \cos \theta, k \sin \theta)$ crosses over between the expected $k^{-2}$ and $k^{-4}$ behaviors, as seen in Fig.\ \ref{fig_2-4PSD_add}. Again, from this point of view, $\theta=0,\pi/2$ allow us to elucidate the two existing independent exponents in the most unambiguous way minimizing 
finite-size effects but do not play a more fundamental role.

\begin{figure}[t!]
\centering
\epsfig{clip=,width=0.5\textwidth,file=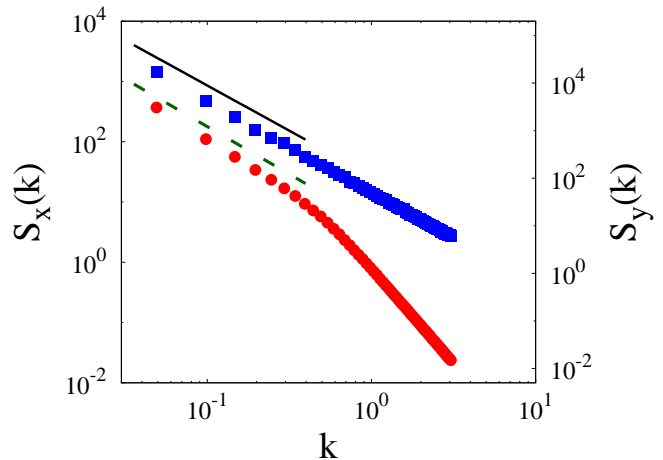}
\caption{(Color online) PSD of one-dimensional cuts $S_x(k)$ (blue squares, left axis) and $S_y(k)$
(red circles, right axis) of surfaces generated by the 2-4 equation \eqref{eq_2-4}. The numerical
integrations were performed with the same parameters as in Fig.\ \ref{fig_2-4PSD}. The solid black line is a guide for the eye with slope $-3/2$, while the dashed green line has slope $-2$ and the dot-dashed magenta line has slope $-4$. All units are arbitrary.}
\label{fig_2-4PSD1d}
\end{figure}

Further, we have also checked that the one-dimensional PSDs of cuts of the surface along the $x$ and $y$ directions are consistent with Eq.\ \eqref{eq_2-4PSD1d}. In Fig.\ \ref{fig_2-4PSD1d} we see that indeed $S_x(k)$ (left axis) and $S_y(k)$ (right axis) display scaling behavior as predicted by the anisotropic Ansatz.
Notice that, for small distances at which dynamics has not yet crossed over to its large-scale behavior, $S_y(k_y) \sim k_y^{-4}$, as would correspond for a 1D profile of a 1D LMBE system \cite{barabasi:book}.

\begin{figure}[t!]
\centering
\epsfig{clip=,width=0.5\textwidth,file=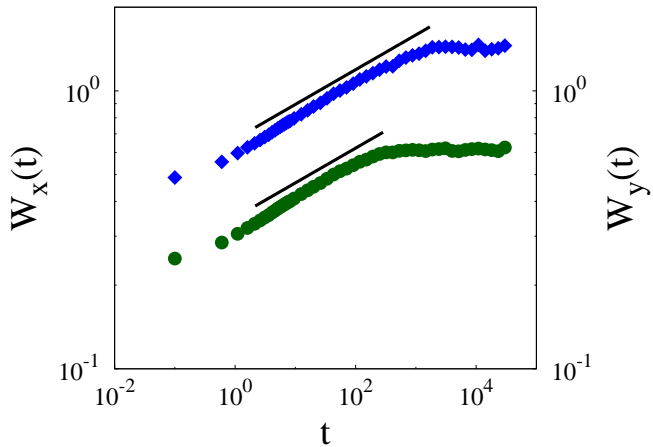}
\caption{(Color online) Time evolution of the surface roughness functions measured along the two spatial directions, $W_x(t)$ (green circles, left axis) and $W_y(t)$ (blue diamonds, right axis), obtained from numerical integrations of the 2-4 equation \eqref{eq_2-4}. For visualization purposes, the value of $W_y(t)$ has been artificially offset vertically. The two solid black lines have the same slopes, $1/8$. Parameters for the simulations are as in \mbox{Fig.\ \ref{fig_2-4PSD}}.
All units are arbitrary.}
\label{fig_2-4W}
\end{figure}

Finally, from the full dynamic evolution of the equation we obtained the value of the growth exponent $\beta$. We have checked that this exponent does not depend on the direction chosen for its estimation.
In Fig.\  \ref{fig_2-4W} we show the  roughness of the surface measured along each space direction.
As expected, 
$W_x(t) \sim W_y(t) \sim t^{1/8}$ for times before saturation.

In summary, for the 2-4 equation, all the anisotropic scaling relations derived in the previous section are verified both numerically and analytically, thus providing a simple example of a system with strong anisotropy. However, it would be interesting to probe these properties in a nontrivial system in which nonlinearities play a prominent role. The aim of the next section is to perform such a study for the celebrated example of the Hwa-Kardar equation.

\section{Nonlinear systems: running sand piles}
\label{sec_hk}

The Hwa-Kardar equation (HK) was originally proposed in the context of self-organized criticality~\cite{hk}.
The intent was to construct a continuum field representation of the \emph{sand-pile} model proposed by Bak, Tang, and Wiesenfeld (BTW) \cite{christensen:book,dickman:2000} and to study its critical behavior within such a ``hydrodynamic'' formulation.

In the BTW model, sand grains are added randomly on a square lattice. At each lattice site, a discrete function $h$ keeps
tracks of the height of the local sand column. Whenever the local height value crosses a threshold, the sand grains of the columns are distributed among the nearest neighbors, until $h$ becomes smaller than the threshold value. If one of the neighboring sites in turn crosses the threshold due to the grains received from the first site, the phenomenon propagates and triggers an \emph{avalanche}. When an avalanche reaches the border of the lattice, the exceeding grains fall out of the system. Many versions of the this model can be found in the literature \cite{christensen:book,dickman:2000}. In the original BTW model, the addition of each grain took place only when the system was completely in a relaxed state (without avalanches). Thus the addition of the grains (driving) and their transport (relaxation) occur at widely separated time scales. In Ref.\ \cite{hk}, the authors focused on the opposite limit of a \emph{running} sand-pile model, in which sand grains are added according to an external clock that is 
independent of the state of the system.

Since transport of sand grains occurs only in the direction parallel to gravity, the resulting field equation is inherently anisotropic. Moreover, due to the fact that the relaxation mechanism implies a global loss of potential energy and the grains are allowed to escape through the boundaries without conservation constraints, the running sand-pile model is dissipative and open. The evolution equation is constructed taking into account the properties of the discrete model  \cite{hk,kardar:book}: The steady state is assumed to be on average a flat surface, $h({\bf r}, t)$ measuring the local deviation from it. Here $r_{\parallel}$ and $\bf r_{\perp}$ are the components of $\bf r$ parallel and perpendicular to the transport direction, respectively. Because of the existence of a preferred flow direction, the system is not invariant with respect to the reflection symmetry in $r_{\parallel}$ or $h$. However, it is assumed to be rotationally invariant in the $\bf r_{\perp}$ direction, translationally invariant in 
both directions, and also invariant under the combined reflections    $h \to -h$ and $r_{\parallel} \to -r_{\parallel}$ \cite{hk,kardar:book}.
Taking into account all the model symmetries, and including the leading order nonlinearity, one obtains 
\begin{equation}
\label{eq_hkd}
\partial_t h = \nu_{\parallel}\partial_{\parallel}^2 h + \nu_{\perp}\nabla_{\perp}^2 h - \frac{\lambda}{2}\partial_{\parallel}h^2 + \eta.
\end{equation}
The first linear terms model the relaxation of the height through diffusive transport, whereas the nonlinearity accounts for the lack of inversion symmetry and is related to the presence of an external driving. In the absence of noise,
dynamics is conserved, namely, Eq.\ \eqref{eq_hkd} takes the form of a continuity equation,
\begin{equation}
\partial_t h + \nabla \cdot {\bf J} = 0,
\end{equation}
where
\begin{equation}
\label{eq_hkc}
{\bf J} = - \nu_{\perp}\nabla_{\perp} h -\left( \nu_{\parallel} \partial_{\parallel}h  -\dfrac{\lambda}{2}h^2 \right)  \hat {\bf t} .
\end{equation}
Here $\hat {\bf t}$ denotes the transport direction along the parallel coordinate. Given that the noise term $\eta$ mimics the random addition of sand particles from outside the system, it is nonconserved, leading to GSI properties
for the HK equation \cite{grinstein:1995}. Here, we will interpret these properties within our anisotropic scaling
Ansatz, specifically for the HK equation in two dimensions ($d=2$),
\begin{equation}
\label{eq_hk2}
\partial_t h = \nu_x \partial_x^2 h + \nu_y \partial_y^2 h - \dfrac{\lambda}{2}\partial_x h^2 + \eta,
\end{equation}
in which the $x$ coordinate is parallel to the transport direction and the $y$ coordinate is perpendicular to it.
Although Eq.\ \eqref{eq_hk2} is nonlinear and cannot be solved analytically, it is a very special system for
which the three scaling exponents characterizing its SA behavior are believed to be exactly known, due to the occurrence of large number of symmetries \cite{hk,kardar:book}. Specifically, these lead to
\begin{eqnarray}
&& z_x = 2\zeta, \label{rel_01}\\[3pt]
&& z_x+\alpha_x = 1, \label{rel_02}\\[3pt]
&& z_x = 2 \alpha_x + \zeta + 1. \label{rel_03}
\end{eqnarray}
\begin{figure}[t!]
\epsfig{clip=,width=0.7\linewidth,file=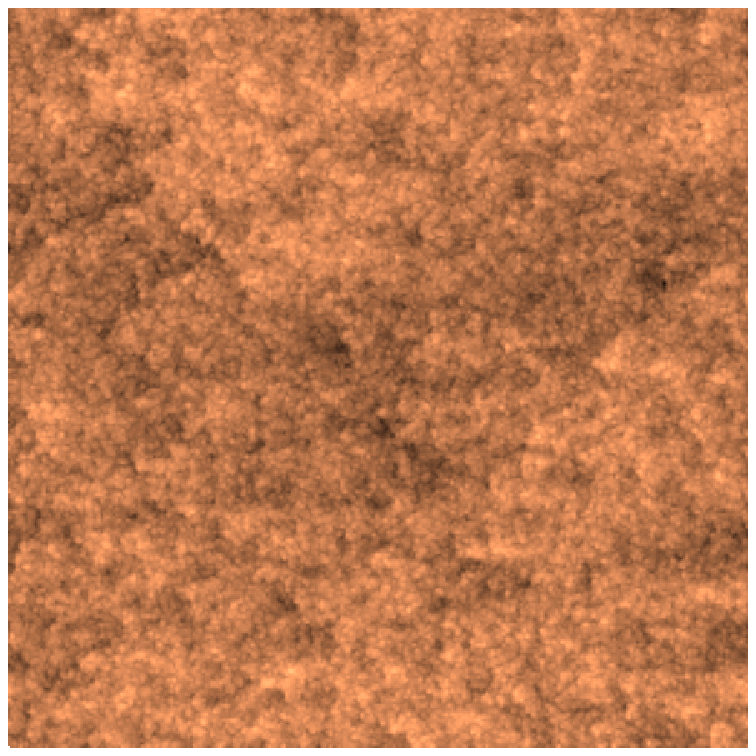}
\vskip 1cm
\epsfig{clip=,width=0.7\linewidth,file=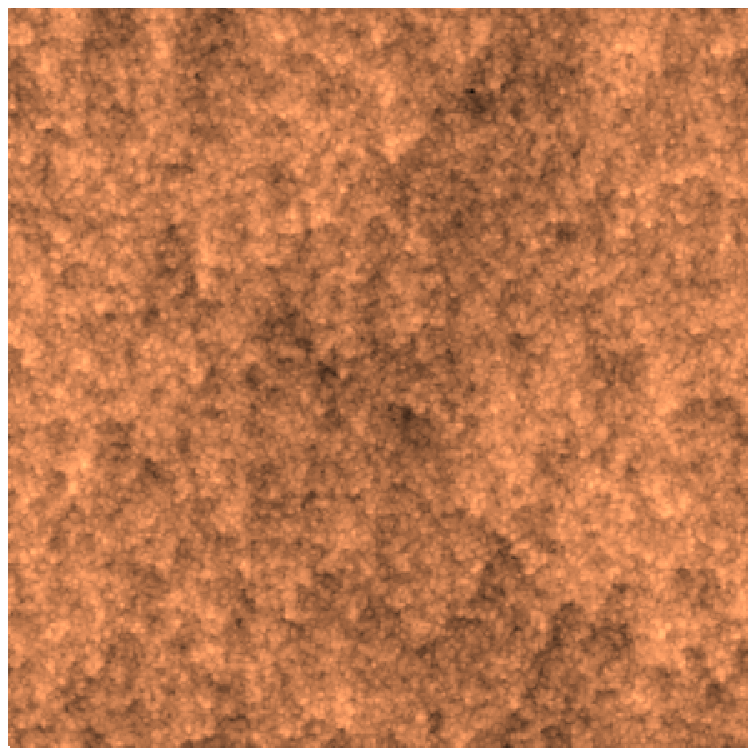}
\caption{(Color online) Top views of the morphology for  surfaces generated by use of the two-dimensional HK equation \eqref{eq_hk2}. Numerical simulations have been performed with parameters $\nu_x = \nu_y = 1$, $\lambda = 3$, $D=1$, $L=256$, $\Delta x = 1$, and $\Delta t = 10^{-2}$. The bottom panel has been obtained after rotating the top panel by $90^{\circ}$.}
\label{fig:morph_hk}
\end{figure}
Equation \eqref{rel_01} originates from the fact that the nonlinearity $\lambda$, acting only along the $x$ direction, cannot
renormalize linear operators that, like the one with coefficient $\nu_y$, act along the perpendicular direction. Equation \eqref{rel_02} implements a vertex cancellation that is believed to hold to arbitrary order in a perturbative
expansion due to a concomitant symmetry of the equation under a Galilean-type transformation parametrized by $\lambda$.
Finally, Eq.\ \eqref{rel_03} is a hyperscaling relation (for $d=2$) \cite{barabasi:book} deriving from the fact
that Eq.\ \eqref{eq_hk2} implements conserved dynamics with nonconserved noise. As mentioned, the three scaling laws
Eqs.\ \eqref{rel_01}-\eqref{rel_03} are believed to hold exactly for arbitrary substrate dimension; see Refs.\ \cite{hk,kardar:book}. For each value of $d$, such a nonhomogeneous linear system actually has as a unique solution the set
of three independent exponents that fully characterize the ensuing SA behavior. In our 2D case, the values are
\begin{equation}
\label{eq_hk_2exp}
\alpha_x = -\dfrac{1}{5}, \quad z_x = \dfrac{6}{5}, \quad \zeta = \dfrac{3}{5}.
\end{equation}
From the relation $z_x = \alpha_x/\beta$, we get $\beta  = -1/6$. Since the growth exponent is negative, subleading corrections appear in the time evolution of the roughness that complicate the estimation of $\beta$ from the study of, e.g., $W_x(t)$ and $W_y(t)$. However, the scaling relations derived in Sec.\ \ref{sec_ansatz} allow us to to compute
\begin{equation}
\label{eq_hk2_expy}
\alpha_y = -\frac{1}{3}, \qquad z_y = 2,
\end{equation}
and the values of the momentum-space exponents of the two-dimensional PSD $S(k_x,k_y)$,
\begin{equation}
\label{eq_hk_exp2d}
\tilde \alpha_x = \frac{3}{5}, \qquad \tilde \alpha_y = 1.
\end{equation}
These will be simpler to check in simulations.
The fact that all the critical exponent values can be calculated analytically makes
the HK equation a perfect case study in which to check the scaling Ansatz \eqref{eq_ASA} as a paradigm
of a nonlinear system displaying strong anisotropy. 
Moreover, to the best of our knowledge, to date there is no available study of the 2D HK equation through direct numerical simulations, in which the anisotropic scaling is analyzed.

\begin{figure}[t!]
\centering
\epsfig{clip=,width=0.5\textwidth,file=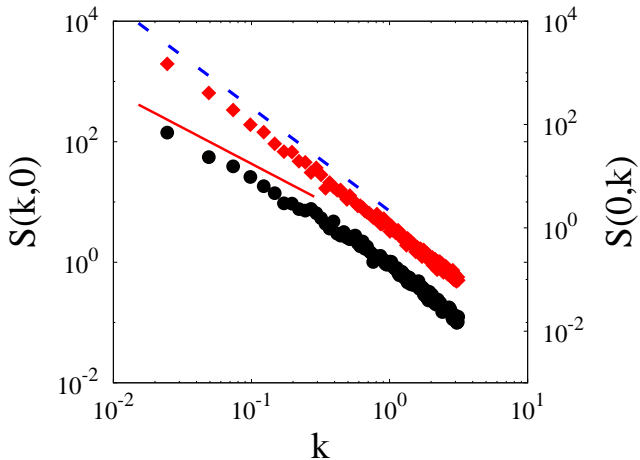}
\caption{(Color online) One-dimensional projections $S(k,0)$ (black circles, left axis) and $S(0,k)$ (red diamonds, right axis) of the two-dimensional PSD for surfaces generated by the two-dimensional HK equation \eqref{eq_hk2}. Numerical simulations have been performed with parameters as in Fig.\ \ref{fig:morph_hk}. The saturated two-dimensional PSD was averaged over 100 different realizations. The solid red line is a guide for the eye with slope $-6/5$, while the dashed blue line has slope $-2$. All units are arbitrary.}
\label{fig_hkPSD}
\end{figure}

As for the 2-4 equation, the scaling behavior of the 2D HK equation \eqref{eq_hk2} has been
verified through pseudospectral numerical simulations. An example of the surface morphology that is
obtained for long simulation times is provided in Fig.\ \ref{fig:morph_hk}.
In order to get a visual impression on the anisotropy of the system, in the same figure we present the same topography
after a $90^{\circ}$ rotation in the $(x,y)$ plane. More quantitatively, in Fig.\ \ref{fig_hkPSD} we present
the two projections of the two-dimensional PSD $S(k,0)$ (left axis) and $S(0,k)$ (right axis).
Again, we clearly observe a scaling behavior like the one predicted by our anisotropic Ansatz, using the exact values Eq.\ \eqref{eq_hk_exp2d}. 

\begin{figure}[t!]
\centering
\epsfig{clip=,width=0.5\textwidth,file=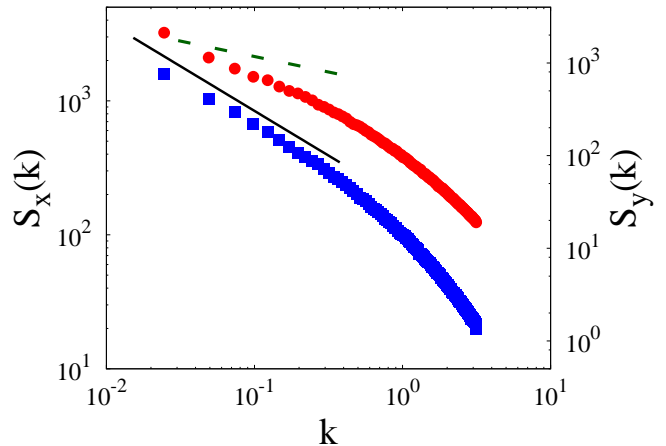}
\caption{(Color online) PSD of one-dimensional cuts $S_x(k)$ (blue squares, left axis) and $S_y(k)$
(red circles, right axis) of surfaces generated by the two-dimensional HK equation \eqref{eq_hk2}. Numerical
integrations were performed with the same parameters as in Fig.\ \ref{fig_hkPSD}. The solid black line is a guide for the eye with slope $-2/3$ while the dashed green line has slope $-1/3$. Note that $S_y(k)$ does {\em not} follow the predicted scaling behavior at small $k$. All units are arbitrary.}
\label{fig_hkPSD1d}
\end{figure}

However, for the PSDs of 1D cuts of the surface, we find a striking difference. As can be seen in Fig.\ \ref{fig_hkPSD1d},
while the long-distance behavior of $S_x(k_x)$ is as expected, this is not the case for the PSD of the one-dimensional cuts along the $y$ direction, $S_y(k)$, that shows an unexpected behavior in the region of small wave vectors.
The discrepancy between the numerical data and our theoretical prediction consists of two different effects: {\em (i)} The value of the PSD for the smallest value of $k$, i.e., $S_y(2\pi/L)$, behaves quite differently with respect to the other points in the curve, ``spoiling'' the scaling behavior. Further numerical inspection (not shown) demonstrates that this feature is not due to fluctuations or numerical artifacts. {\em (ii)} Even neglecting this point, the scaling behavior of $S_y(k)$ does not resemble in any way the theoretical one given by Eqs.\ \eqref{eq_hk2_expy}. This unexpected behavior of $S_y(k)$ deserves to be studied with more detail since it could have important consequences about the anisotropic scaling  Ansatz \eqref{eq_ASA}. As potential explanations, we can conceive of three possibilities: The Ansatz needs modifications in the presence of nonlinearities,  the particular values of the exponents induce this behavior, or both the nonlinearity and the exponent values concur in 
producing it. In order to elucidate this phenomenon, in the next section
we construct a linear equation with the same scaling exponents of the HK equation. This nonlocal Gaussian approximation can be solved analytically and satisfies the anisotropic scaling Ansatz.

\subsection{(Non-local) Gaussian approximation}
\label{sec_gauss}

The fact that the HK asymptotics is, in principle, determined by three scaling relations allows us to put forward
a {\em linear} equation that shares the same three independent exponents exactly. The key observation appears already
in Eq.\ \eqref{ec_resc} of the previous section. Namely, for an $(n,m)$ equation, a rescaling allows us to fix all three
exponents by requesting statistical invariance. In order to see it,
let us rescale Eq.\ \eqref{eq_nm} for $n=2\tilde\alpha_x$, $m=2\tilde\alpha_y$, and $d=2$, according to the transformation $h' = b^{\alpha_x}h$, $k'_x =b^{-1}k_x$, $k'_y=b^{-\zeta}k_y$, and $t'=b^{z_x} t$ [which is the momentum-space counterpart  of $\vec{r}_1$, see Eq.\ \eqref{r1}]. Requesting scale invariance, we obtain three relations among the critical
exponents, namely
\begin{eqnarray}
&& 2\tilde\alpha_x = z_x , \label{rel_1}\\[3pt]
&& 2\zeta\tilde\alpha_y = z_x ,\\[3pt]
&& z_x = 2\alpha_x + \zeta +1 . \label{rel_3}
\end{eqnarray}
From Eqs.\ \eqref{eq_relalphax}, \eqref{eq_relalphay}, and \eqref{eq_eqbeta}, it is straightforward to see that the
three relations \eqref{rel_1}-\eqref{rel_3} are completely equivalent, so  only one of them is independent. Actually, Eq.\ \eqref{rel_3} is {\sl hyperscaling}, which holds for any linear equation and is also shared by the HK equation.
For a nonconserved nonlinearity (for example, of the KPZ type), the noise term becomes renormalized and a such constraint may not hold. In such a case, the nonlinear equation and its Gaussian approximation would not share the same set of scaling exponents, in contrast with the case considered here.

After these preliminary considerations, it is straightforward to find a $(n,m)$ equation that has the
same three independent exponents as the nonlinear HK equation. Thus, we just need to take $n = 2\tilde\alpha_x = 6/5$ and
$m =2 \tilde\alpha_y = 2$. That is, the equation
\begin{equation}
\label{eq_hkG}
\partial_t \tilde{h}({\bf k},t) = -\left( \nu_x |k_x|^{6/5} + \nu_y |k_y|^2\right) \tilde{h}({\bf k},t)
+ \tilde{\eta}({\bf k},t),
\end{equation}
has, by construction, the exponents given in \eqref{eq_hk_2exp}. Note that this equation is non-local
in space since $n=6/5$ is not an even integer. Moreover, it is important to stress that Eq.\  \eqref{eq_hkG} is {\em not} the linearized HK equation. The latter would be obtained by setting $\lambda$ equal to zero, and, therefore, both exponents in momentum space would be $2\tilde\alpha_x=2\tilde\alpha_y=2$, corresponding to the simple two-dimensional Edwards-Wilkinson equation \cite{barabasi:book}. On the other hand, Eq.\ \eqref{eq_hkG} can be considered  a Gaussian approximation
of the HK equation, in the sense that it is a variational equation \cite{kardar:book} that has the asymptotic
Boltzmann height distribution ${\cal P}\{h\} \propto \exp[-(1/2D) \int \left( \nu_x |k_x|^{6/5} + \nu_y |k_y|^2\right) |\tilde{h}({\bf k})|^2 {\rm d}\mathbf{k}]$. The fact that accurate approximations of nonlinear stochastic equations can be achieved through appropriate Gaussian counterparts is already known. Thus, in equilibrium, the Bogoliubov-Feynman's inequality leads to a variational mean-field approximation that can be applied even to systems for which nonlinearities
are non polynomial in the height field \cite{saito:book,moro:2002}, while, out of equilibrium, the so-called self-consistent
expansion (see Ref.\ \cite{schwartz:2008} for an overview with references) allows us to account for the scaling exponents of a number of-nontrivial systems through an appropriate Gaussian Ansatz. 

\begin{figure}[t!]
\centering
\epsfig{clip=,width=0.5\textwidth,file=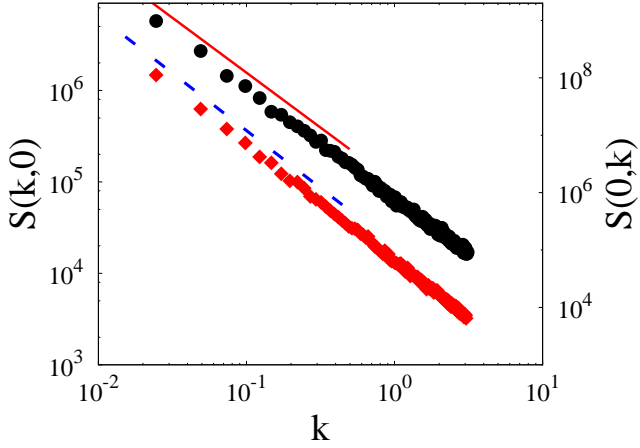}
\caption{(Color online) One-dimensional projections $S(k,0)$ (black circles, left axis) and $S(0,k)$
(red diamonds, right axis) of the two-dimensional PSD of surfaces generated by the Gaussian approximation of the
HK equation, Eq.\ \eqref{eq_hkG}. Numerical integrations have been performed with parameters $\nu_x = \nu_y = 1$, $D=1$, $L=256$, $\Delta x = 1$, and $\Delta t = 0.1$. The saturated two-dimensional PSD has been averaged over 300 different realizations. The solid red line is a guide for the eye with slope $-6/5$ while the dashed blue line has slope $-2$.
All units are arbitrary.}
\label{fig_hkGPSD}
\end{figure}

Since all the scaling relations that we have derived stem from the two-dimensional PSD, for Eq.\ \eqref{eq_hkG} we expect the behavior of $S_x(k)$ and $S_y(k)$ to be exactly as predicted by \eqref{eq_sSx} and \eqref{eq_sSy}. Before analyzing the PSD of one-dimensional cuts of the surface, we have checked the behavior of the two-dimensional PSD. In Fig.\ \ref{fig_hkGPSD} we plot the projections of the 2D PSD  onto the $k_x$ and $k_y$ axes, as in previous cases.
The scaling for small $k$ is as expected.

\begin{figure}[t!]
\centering
\epsfig{clip=,width=0.5\textwidth,file=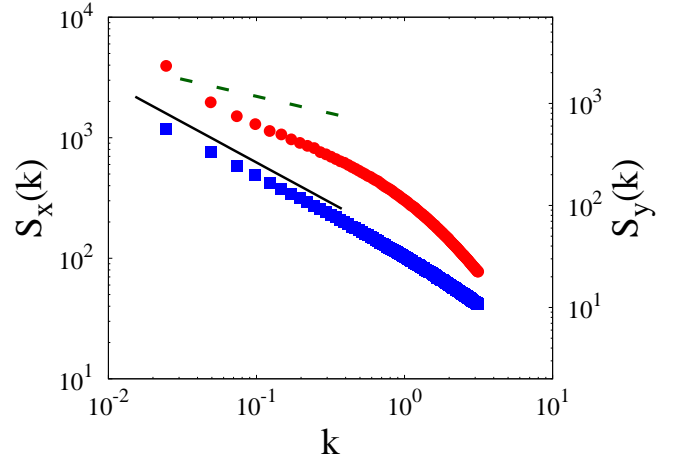}
\caption{(Color online) PSD of one-dimensional cuts $S_x(k)$ (blue squares, left axis) and $S_y(k)$ (red circles, right axis) of surfaces generated by the Gaussian approximation of the HK equation [Eq.\ \eqref{eq_hkG}]. The numerical integrations have been performed with the same parameters as in Fig.\ \ref{fig_hkGPSD}. The solid black line is a guide for the eye with slope $-2/3$ while the dashed green line has slope $-1/3$. Note that $S_y(k)$ does {\em not} follow the predicted scaling behavior. All units are arbitrary.}
\label{fig_hkGPSD1d}
\end{figure}

Moreover, $S_x(k)$, reported in Fig.\ \ref{fig_hkGPSD1d} (left axis), behaves as expected.
However, the one-dimensional PSD of cuts along the $y$ axis $S_y(k)$ (right axis) displays the same unexpected behavior as for the nonlinear HK equation, the first point being ``shifted'' above the rest of them and not adjusting
to a straight line with the expected slope. Thus, such  behavior does not seem due to the scaling Ansatz
we are employing since, for a linear equation like \eqref{eq_hkG}, this simply expresses the exact behavior of the observables in the long time limit. Thus, the behavior of $S_y(k)$ seems, rather, induced by the present exponent values.

In order to clarify this issue, it is important to keep in mind that we are working for a finite system of lateral size $L<\infty$, and with a finite lattice spacing $a=\Delta x\neq 0$.
The discrete counterparts of Eqs.\ \eqref{eq_Sx} and \eqref{eq_Sy} are
\begin{eqnarray}
&& \hat{S}_x(k_x) = \dfrac{1}{L}\left[\hat{S}(k_x,0) + 2\sum_{n_y=1}^{N/2}\hat{S}(k_x,k_y) \right],\\
&& \hat{S}_y(k_y) = \dfrac{1}{L}\left[\hat{S}(0,k_y) + 2\sum_{n_x=1}^{N/2}\hat{S}(k_x,k_y) \right], \label{eq_discSy}
\end{eqnarray}
where $N=L/\Delta x=L/a$, and $n_x,n_y=1,\dots,N$ are discrete indices so  $k_{x,y} =2\pi n_{x,y}/L$.
The hat stresses the discrete nature of these formulas. In the thermodynamic limit (for $L\to \infty$), these two series can be well approximated as
\begin{eqnarray}
&& \hat{S}_x(k_x) \sim\dfrac{1}{\pi}\int_0^{\pi/a}dk_y \, S(k_x,k_y)\nonumber \\[5pt]
&& \hskip 40pt =k_x^{-(2\tilde\alpha_x-\zeta)} I(k_x^{\zeta},\tilde\alpha_x),\label{eq_Sxd} \\
&& \hat{S}_y(k_y) \sim \dfrac{1}{\pi}\int_0^{\pi/a}dk_x \, S(k_x,k_y) \nonumber \\[5pt]
&& \hskip 40pt =k_y^{-(2\tilde\alpha_y-1/\zeta)} I(k_y^{1/\zeta},\tilde\alpha_y),\label{eq_Syd}
\end{eqnarray}
where we have defined
\begin{equation}
I(k,\alpha) = \dfrac{1}{\pi}\int_0^{\pi/ak} \dfrac{du}{1+u^{2\alpha}}.
\label{eq_I}
\end{equation}
In case the integral $I$ is finite in the continuum $a\to 0$ limit, the approximations \eqref{eq_Sxd} and \eqref{eq_Syd} hold and we indeed obtain the scaling laws \eqref{eq_sSx} and \eqref{eq_sSy}. However, {\sl depending on the value of the
critical exponents} and the numerical parameters employed, convergence of the integrals in \eqref{eq_Sxd} and \eqref{eq_Syd} can be too slow, and a clear-cut scaling may be jeopardized.

\begin{figure}[t!]
\centering
\epsfig{clip=,width=0.5\textwidth,file=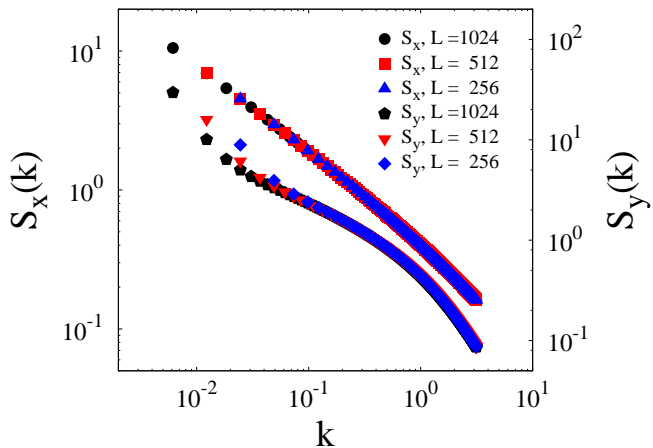}
\caption{
(Color online) Discrete 1D PSDs $\hat{S}_x(k)$ (left axis) and $\hat{S}_y(k)$ (right axis) calculated from the sampled two-dimensional PSD $\hat{S}(k_x,k_y)$ of the Gaussian approximation Eq.\ \eqref{eq_hkG} for increasing system size $L$.
Other model parameters are the same as in Fig.\ \ref{fig_hkGPSD}. All units are arbitrary.}
\label{fig_discS}
\end{figure}

\begin{figure}[t!]
\centering
\epsfig{clip=,width=0.5\textwidth,file=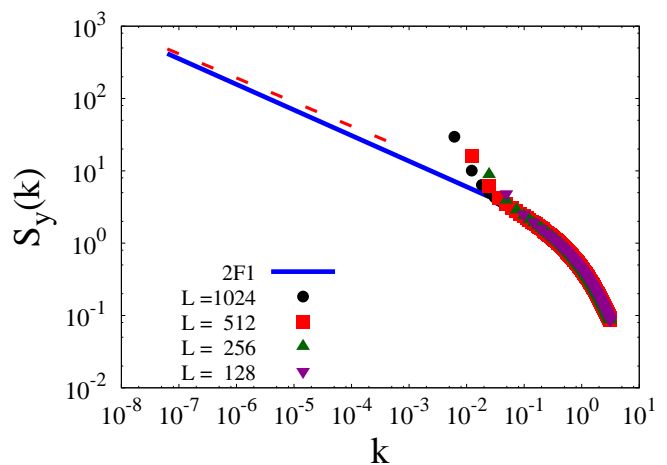}
\caption{(Color online) Comparison of simulation data for $S_y(k)$ as obtained for Eq.\ \eqref{eq_hkG} in Fig.\ \ref{fig_discS}, with the analytical solution of Eq.\ \eqref{eq_discSy} given by Eq.\ \eqref{eq_F} (solid blue line). The dashed red line
corresponds to the asymptotic behavior $S_y(k) \sim k^{-1/3}$. All units are arbitrary.}
\label{fig_exactSy}
\end{figure}

In order to verify this issue, we have evaluated the series $\hat{S}_x(k)$ and $\hat{S}_y(k)$
from the known two-dimensional PSD $\hat{S}(k_x,k_y)$ for increasing system size $L$ (at fixed space
resolution $a$). In Fig.\ \ref{fig_discS} we see that the asymptotic scaling is already clear-cut for $\hat{S}_x(k)$ for
the $L$ values considered, while for $S_y(k)$ it is still far from being seen.
However, in the figure one can see that the curves for $\hat{S}_y(k)$ become smoother as $L$ increases, supporting the prediction that, for a sufficiently large system size, a well-defined scaling behavior could be seen. A further step can be taken by solving the integral in Eq.\ \eqref{eq_Syd} and \eqref{eq_I} exactly,
\begin{equation}
\lim_{L\to \infty} \hat{S}_y(k_y)\sim k_y^{-2\tilde\alpha_y}\, a^{-1}\, {}_2F_1\left(p_1, 1;1+p_1,p_2\right),  \label{eq_F}
\end{equation}
where ${}_2F_1$ is a hypergeometric function with parameters $p_1=1/2\tilde\alpha_x$ and
$p_2 = -(\pi/a)^{2\tilde\alpha_x} k_y^{-2\tilde\alpha_y}$. In Fig.\ \ref{fig_exactSy} we show this approximation of $\hat{S}_y(k_y)$ for a huge system ($L=10^8$) with lattice cutoff $a=1$.
For the largest system sizes, the asymptotic scaling is clear and agrees perfectly with the one predicted from our
scaling Ansatz. In the figure we, moreover, compare the analytical behavior with the numerical data from simulations of
different system sizes. The numerical data indeed tend to adjust to the theoretical curve. This proves that, for the present values of the exponents, the sum \eqref{eq_discSy} for $S_y(k_y)$ converges too slowly to the integral \eqref{eq_F} for any numerically feasible system size. On the other hand, $S_x(k_x)$ displays the asymptotic behavior already at relatively small values of $L$. Thus, the convergence rate strongly depends on the values of the critical exponents. One way to directly verify this would be to perform numerical simulations with an adequate parameter set, corresponding to a theoretical curve in which the scaling behavior is clear. Unfortunately, as suggested by Fig.\ \ref{fig_exactSy}, this would require a  computational effort beyond our capabilities.

These results for the Gaussian approximation of the HK equation support the hypothesis of a similar slow-convergence problem also for the nonlinear HK equation proper. The fact that the behavior of $S_y(k_y)$ for Eqs.\ \eqref{eq_hk2} and \eqref{eq_hkG} is qualitatively the same, and the fact that the presence of a nonlinearity usually contributes to slowing the convergence rate of observables, lead us to think that the asymptotic state for HK equation has not been fully reached in our simulations. However, as for its Gaussian approximation, we have not been able to perform a sufficiently large-scale numerical simulation due to the great computational effort required.

The present slow-convergence problem may have important implications for the interpretation of experimental data.
As pointed out earlier, many experimental setups are designed to measure the PSD of one-dimensional cuts, but we have shown here that for such integrated quantities there are systems in which the scaling may not be clearly detected. On the other hand, the two-dimensional PSD appears to be a more robust quantity, since it scales correctly already for relatively small values of the system size $L$ and relatively large values of the spatial discretization. We have recently confirmed
these expectations in the analysis of experiments on surface erosion by ion-beam sputtering \cite{us_exp}.

\section{Discussion and conclusions}
\label{sec_disc}


In this work we have presented a dynamic scaling Ansatz for two-dimensional anisotropic kinetic roughening that has a form readily applicable to the analysis of simulation or experimental data. The Ansatz incorporates the behavior that characterizes strongly anisotropic linear equations, as has been thoroughly checked via their exact solution and numerical simulations. In this process, we have introduced a family of linear models that display SA while generally being nonlocal in space. Incidentally, nonlocal models (incorporating, e.g., fractional Laplacians) are lately finding ubiquitous use in equilibrium and non-equilibrium statistical mechanics \cite{henkel:book_v2}; see Refs.\ \cite{nicoli:2009a,nicoli:2009b,nicoli:2011} and references therein for the specific context of kinetic roughening.

A natural second step that we have taken in the present paper has been the validation of our scaling Ansatz
against a numerical study of a paradigmatic nonlinear model of SA and in general of GSI systems, namely
the HK equation for running sand piles. Again, agreement is quite satisfactory (as occurs for some
experimental systems \cite{us_exp}) and can be concluded after elucidation of slow convergence properties of certain observables, induced by (small) critical exponent values. This has required us to formulate and study a linear equation that has the same exact exponents as the HK nonlinear system and, thus, can be considered  a Gaussian approximation of the latter. Thus, our results for a representative nontrivial anisotropic system suggest that the scarcity in assessments of anisotropic scaling properties in practical cases may be related with the occurrence of similar finite size-effects.

Methodologically, our present work provides some results that can be of interest in the analysis of strongly anisotropic
interfaces. On the one hand, we have seen that in the case of systems with conserved dynamics and nonconserved noise
it is always possible to tailor a fully dynamical description through an appropriate $(n,m)$ linear equation. This is due to the fact that the ensuing hyperscaling relation reduces the number of independent exponents to two that are then readily
tuned by choosing the {\em real} (non-necessarily integer) parameters $n, m$ appropriately. Such a procedure to put forward an effective dynamical model may be of interest if dealing with complex nonlinear systems and/or with the description of simulation and/or experimental data. Incidentally, for nonconserved dynamics, some examples are also known on the equivalent stationary-state properties of some nonlinear models ---such as the anisotropic KPZ equation for nonlinearities with opposite signs--- and of linear equations \cite{dasilveira:2003}. Likewise, a whole family on nonlocal and nonlinear models exists which shares the same probability distribution function at steady state with nonlocal linear models akin to the present $(n,m)$ equations; see Ref.\ \cite{katzav:2003}.

On the other hand, our analysis has shown the practical convenience of the use of certain observables (e.g., 2D
PSD function) over others (e.g., PSD functions of 1D cuts of the 2D surface), provided one can access  data
with a large-enough signal-to-noise ratio. This has actually been recently assessed in the analysis of experimental data
from thin-film production \cite{us_exp}.

In order to go beyond the present paper, one possibility would be to consider the applicability of the exponent inequality that has been recently found for large classes of growth models in Ref.\ \cite{katzav:2011}, which leads to estimates of the scaling exponents even in nonlinear systems. This would require a generalization of the behavior of the various observables involved (response and correlation functions) in the anisotropic case. Another natural avenue to push forward our present results is to consider the pertinence of our scaling Ansatz to nonlinear equations whose dynamics lies beyond reach of linear (Gaussian) approximations. Questions of interest in this connection are the occurrence, if at all, of SA behavior of the type we have discussed here and under which conditions. Note in particular, that all the systems we have analyzed so far correspond to conserved dynamics. Thus, the case of nonconserved system becomes particularly intriguing, especially in view of the fact that the natural paradigmatic 
system, namely the anisotropic KPZ equation, does {\em not} show strongly anisotropic scaling in two dimensions \cite{wolf:1991}.

\begin{acknowledgments}

We are indebted to A.\ Keller for discussions. Partial support for this work was provided by MICINN (Spain) Grant No.\ FIS2009-12964-C05-01. E.\ V.\ acknowledges support by Universidad Carlos III de Madrid.

\end{acknowledgments}

\end{document}